\documentclass[12pt,a4paper]{article}
\newcommand{\half}{\mbox{${\textstyle \frac{1}{2}}$}}           
\usepackage{graphics}

\begin{document}
\baselineskip 4ex
\vspace*{-2cm}
\hfill{\bf TSL/ISV-96-0162}
\vspace{1.5cm}
\begin{center}
\vspace*{10mm}
{\Large{\bf Comparison of the Near-Threshold Production} }\\[3ex] 
{\Large{\bf of} 
\mbox{\boldmath $\eta$}{\bf - and K-Mesons in Proton-Proton Collisions}}
\\[7ex]
{\large G{\"{o}}ran F\"{a}ldt}\footnote{Electronic address: faldt@tsl.uu.se}
\\[1ex]
{\normalsize Division of Nuclear Physics, Box 533, 751 21 Uppsala, Sweden}
\\[3ex]
{\large Colin Wilkin}\footnote{Electronic address: cw@hep.ucl.ac.uk}\\[1ex]
{\normalsize University College London, London, WC1E 6BT, UK}\\[9ex]
\end{center}

\begin{abstract}
The $pp\to pp\eta$ and $pp\to p\Lambda K^+$ reactions near threshold are 
dominated by the first and second S$_{11}$ resonance respectively. It is shown
that a one-pion-exchange model exciting these isobars reproduces well the ratio
of the production cross sections. The consequences for this and other channels 
are discussed.
\end{abstract}
\vspace{2cm}
PACS numbers: 13.60Le, 13.75Cs, 14.40Aq\\[2ex]
Keywords: threshold meson production, final state interactions

%
%
\newpage

The production of $\eta$ mesons near threshold has been subjected to intense 
experimental effort in the last few years for both nucleon-nucleon
\cite{Ulf,Bing,Beppe} and nucleon-nucleus collisions \cite{Ben}. The low energy 
$\eta$-nucleon interaction is very strong because of the presence of the first 
S$_{11}$ resonance, the N$^*(1535)$, which has a large branching ratio into 
$\eta N$ and whose width overlaps this threshold.

The measurements \cite{Ulf} of the `elementary' $pp\to pp\eta$ 
reaction, shown in fig.~1, exhibit an energy dependence which is dominated by
the proton-proton final state interaction (FSI) weighted by the three-body
phase space. This is easily understood in terms of the one-meson-exchange model
illustrated in fig.~2 \cite{X}. The high momentum transfers necessary for the
production of such a large mass means that the S-wave amplitude depends
primarily on the $pp$ wave function at very short distances, with only a weak
dependence arising from the incident beam energy. It has to be borne in mind
that an excitation energy $Q$ of say 20~MeV is very small compared to the
N$^*(1535)$ width of 150~MeV. The strengths of the different meson exchanges
then merely fix the normalisation of the cross section and there is here
considerable controversy on the importance of the $\pi$, $\rho$ and $\eta$
terms \cite{X}. Their relative size influences significantly the ratio of
$\eta$ production in $pp\to pp\eta$ and $pn\to pn\eta$ collisions, the latter
being more copious by about a factor of six \cite{Beppe,Stina}.

The first measurement of $K^+$ production {\it via} the analogous $pp\to
p\Lambda K^+$ reaction close to threshold ($Q=2$~MeV) has just been reported
\cite{Jan}. It is important to note that the threshold $K^+\Lambda$ system is
dominated by the \underline{second} S$_{11}$ resonance, the N$^*(1650)$. This
means that forms of the production operators and all the spin-angular momentum
algebra are identical in the two cases; the observation of an $\eta$ or a $K$
merely tags which of the two S$_{11}$ resonances has been excited. It is
therefore worthwhile asking if the same type of meson exchange model is capable
of explaining simultaneously the near-threshold $\eta$ and $K$ production data.

The form of the production operator corresponding to the pion-exchange 
contribution to the $pp\to pp\eta$ amplitude shown in fig.~2 is discussed in
detail in ref.\cite{X} but at high momentum transfers it reduces effectively to
a $\delta$-function in the inter-baryon separation vector $\vec{r}$ \cite{FW1},
\begin{equation}
\label{1}
t_{\pi} = \lambda f_{\pi^0p\to \eta p}\,\delta(\vec{r})\:,
\end{equation}
where $f_{\pi^0p\to \eta p}$ is the $\eta$-production amplitude in the 
pion-nucleon channel. The value of $\lambda$ is influenced by the coupling 
constant, form factors, distortion {\it etc}. 

We have previously shown \cite{FW1,FW2} that the S-wave proton-proton wave
function at short distances has an energy dependence of the form
\begin{equation}
\label{2}
\left| \psi_{k}(0)\right|^2 \approx \frac{1}{k^2+\alpha^2}\:,
\end{equation}
where $\vec{k}$ is the proton momentum and $\alpha$ the wave number of the
virtual bound state in the $pp$ system. Taking the parameters of the Paris 
wave function \cite{Paris}, given in table~1, and neglecting the energy 
dependence resulting from the pion-nucleon amplitude, this form gives an 
excellent description of the energy dependence of near-threshold pion 
production in $pp\to pp\pi^0$; the shape is indistinguishable from the
predictions of more microscopic calculations.

In order to compare $\eta$ and $K^+$ production quantitatively, we need not
only the energy dependence of the zero-range baryon-baryon wave function but
also its normalisation. Due to their simplicity, we parametrise the S-wave $pp$
and $\Lambda p$ systems using Bargmann potentials \cite{Newton}. In terms of
$\alpha$ and a shape parameter $\beta$, the scattering length and effective
range are given by
\begin{equation}
\label{3}
a=\frac{\alpha+\beta}{\alpha\beta}\,,\ \ \ \ r=\frac{2}{\alpha+\beta}\:\cdot
\end{equation}
There is much uncertainty in the values of the low energy spin-singlet and
spin-triplet $\Lambda p$ parameters but, taking the theoretical and
experimental estimates given respectively in refs.\cite{param} and
\cite{Gideon} as representative, we find the values of $\alpha$ and $\beta$ 
quoted in table~1.

The non-relativistic three-body phase space weighted with the FSI energy
dependence of eq.(\ref{2}) can be easily integrated to yield cross sections
\begin{eqnarray}
\label{4}
\sigma(pp\rightarrow p \Lambda K^+)&=&C\,Q^2(\half Z_s(Q)+Z_t(Q))
|f_{\pi^0p\to K\Lambda}|^2 \:,\\
\label{5}
\sigma(pp\rightarrow p p \eta)&=&C\,Q^2\, Z(Q)\,|f_{\pi^0p\to \eta p}|^2\:,
\end{eqnarray}
where $C$ is the \underline{same} constant for both $\eta$ and $K$ production,
and separate contributions of the different $\Lambda p$ spin states have been
summed.

The final state interactions are represented by the functions $Z(Q)$, whose
generic expression is
\begin{equation}
\label{6}
Z(Q)=\frac{\beta^2}{\alpha^2}\,\frac{4}{(1+\sqrt{1+Q/\epsilon})^2}\:,
\end{equation}
where $\epsilon=\alpha^2/2\mu$, with $\mu$ being the reduced mass in the $pp$ 
and $p\Lambda$ systems.

The experimental values of the pion-induced $\eta$ and $K$ production
amplitudes at threshold can be deduced from $\pi^-p$ data and are found to be
$|f_{\pi^0p\to \eta p}|^2 = (380\pm 50)~\mu$b/sr \cite{Binnie} and
$|f_{\pi^0p\to K^+ \Lambda}|^2 = (25\pm 5)~\mu$b/sr \cite{Saxon}. 

The prediction of eq.(\ref{5}) for $pp\to pp\eta$, shown in fig.~1, describes
most of the observed energy dependence of the cross section. The deviation of
about a factor of $1.6$ from the curve at high $Q$ could be the result of final
state interactions of the $\eta$ with a proton but this is a small effect
compared to the factor of 30 variation with energy shown in the figure. The
overall theoretical normalisation is uncertain by at least a factor of two due
to the effects of $\rho$ exchange, form factors {\it etc.}, but the values are
not too dissimilar to those shown in fig.~1 \cite{X}. Choosing rather to fix
the scale by letting the curve pass through the lower points of the $\eta$
data, eq.(\ref{4}) then predicts absolute cross sections for the $pp\to
p\Lambda K^+$ reaction. It is very reassuring that the curves corresponding to
different values of the $\Lambda p$ singlet scattering length
\cite{param,Gideon} pass through the error bars of the one measured point
\cite{Jan}. The curves do however diverge further at higher $Q$.

It would easily be possible to include the finite-range effects in the
production operators, but these lead to comparatively small changes in cross
section predictions \cite{FW1}. In view of the uncertainties of the $\rho$
couplings to the two S$_{11}$ isobars and the poorly determined $\Lambda p$
S-wave parameters, it is unrealistic at present to go much further than the
simplistic zero-range approximation given here. However we have at the very
least shown that it is dangerous to model the $pp\to pp\eta$ and $pp\to
p\Lambda K^+$ reactions independently. A more refined treatment would also have
examine the effects of $K^+$-exchange diagrams on $p\Lambda K^+$ production, 
though these are likely to be weak since no exotic $S=+1$ resonances are known.

The approach will be tested further when new data become available at various
energies in the range $1\leq Q\leq 5$~MeV \cite{Jan}. However, from the
experience gained from exciting the first S$_{11}$ resonance through $\eta$
production, the production rates with a neutron target giving 
$pn\to p\Lambda K^0/n\Lambda K^+$ are likely to be larger by a factor of 5--10
compared to those in $pp$ collisions. It is therefore an experimental challenge
to find ways of measuring them. The COSY $pp\to p\Lambda K^+$ experiment could
herald in a generation of near-threshold $\Lambda K^+$ production experiments
on light nuclei which will be sensitive to the S-wave $\Lambda$-nucleus
interaction.\\[3ex]

We are very grateful to J.T.~Balewski and W.~Oelert for valuable discussions
regarding their experiment reported in ref.\cite{Jan}, and also for support
from the IKP-KFA J\"ulich where the work was initiated. This collaboration has
been made possible by the continued financial aid from the Swedish Royal
Academy of Science and one of the authors (CW) would like to thank them and the
The Svedberg Laboratory for their generous hospitality. 

%
%

%
%
\newpage
\begin{table}[p]
\caption[table1]{Parameters of the Bargmann potential and the virtual state
`binding' energies for the $pp$ and $\Lambda p$ systems. The values for the
former are deduced from the results of the Paris potential \cite{Paris}.
The values in the hyperon-nucleon case are obtained (a) from the theoretical 
predictions of the $\Lambda p$ scattering length and effective range 
$a_s=-1.6$~fm, $r_s=1.4$~fm, $a_t=-1.6$~fm, and $r_t=3.2$~fm from 
ref.\cite{param}, and (b) from experimental values of 
$a_s=-1.8$~fm, $r_s=2.8$~fm, $a_t=-1.6$~fm, and $r_t=3.3$~fm from 
ref.\cite{Gideon}}
\vspace{5mm}
\centering
\vspace{2ex}
\begin{tabular}{|c||c|c|c|}
\hline
&&&\\
&$\alpha$&$\beta$&$\epsilon$\\
System&(fm$^{-1}$)&(fm$^{-1}$)&(MeV)\\
&&&\\
\hline\hline
&&&\\
pp singlet& -0.104&0.845&0.45\\
&&&\\
\hline\hline
&&&\\
$\Lambda$p singlet& -0.470&1.90&8.4\\
(theory)&&&\\
&&&\\
\hline
&&&\\
$\Lambda$p triplet& -0.386&1.01&5.7\\
(theory)&&&\\
&&&\\
\hline\hline
&&&\\
$\Lambda$p singlet& -0.367&1.08&5.1\\
(experiment)&&&\\
&&&\\
\hline
&&&\\
$\Lambda$p triplet& -0.383&0.99&5.6\\
(experiment)&&&\\
&&&\\
\hline
\end{tabular}
\end{table}
\vspace{2cm}
%
%
\newpage
\begin{figure} [p]
\begin{center}
\resizebox{\textwidth}{!}{\includegraphics[0cm,3cm][20cm,26cm]{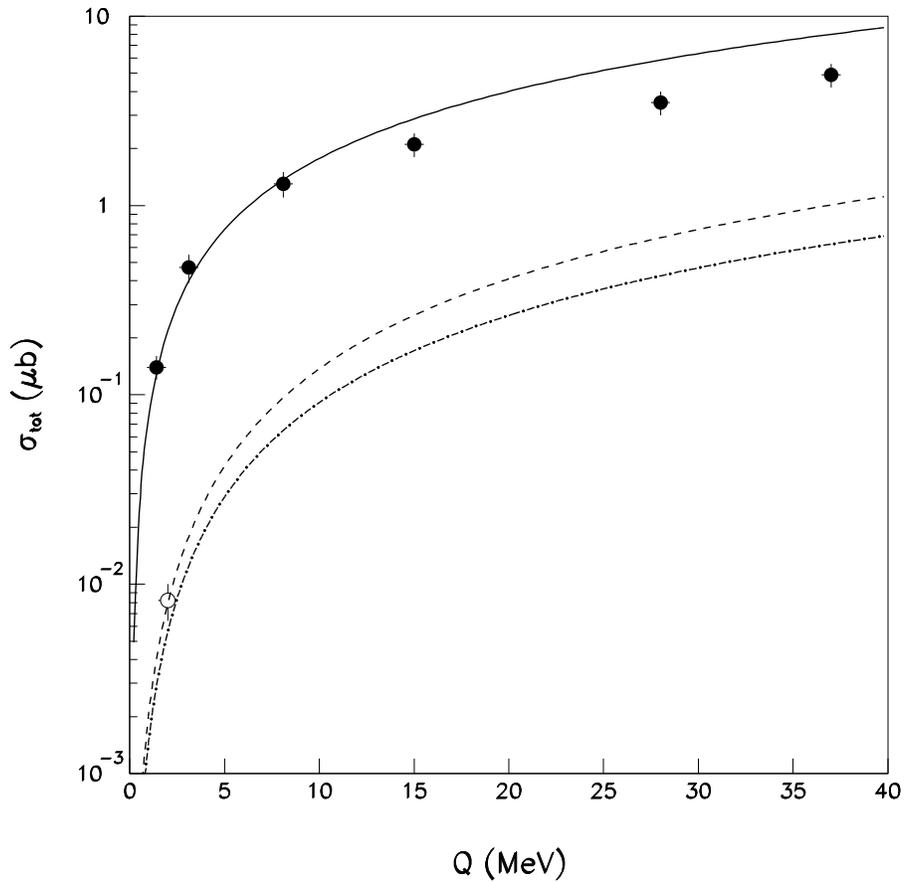}}
\caption{Experimental data for the total cross sections for $pp\to pp\eta$ from
ref.\protect\cite{Ulf} (closed circles) and $pp\to p\Lambda K^+$ from 
ref.\protect\cite{Jan}
(open circle) as functions of $Q$, the kinetic energy in the final state. The
theoretical predictions of eqs.(4,5) are normalised so that the curve passes
through the low energy $\eta$-production points. The broken curve corresponds
to  $\Lambda p$ scattering parameters taken from ref.\protect\cite{param}, 
whereas the
chain curve follows from the experimental parameters of 
ref.\protect\cite{Gideon}.}
\end{center}
\end{figure}%
%
\newpage
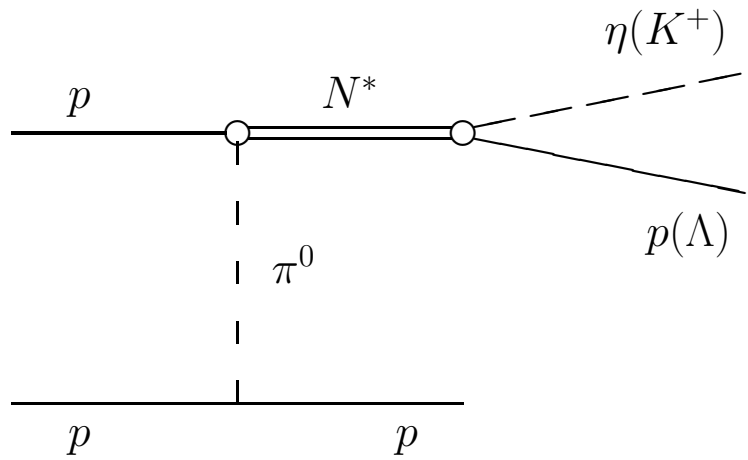
\begin{figure}[p]
\setlength{\unitlength}{1.5mm}
\begin{picture}(140,40)(20,0)
\thicklines
\put(40,6){\line(1,0){40}}
\put(40,30){\line(1,0){19}}
\put(61,30.5){\line(1,0){18}}
\put(61,29.5){\line(1,0){18}}
\multiput(60,6)(0,5.3){5}{\line(0,1){2.0}}
\multiput(81,30.5)(5,1){5}{\line(5,1){3.6}}
\put(81,29.5){\line(5,-1){24}}
\put(60,30){\circle{2}}
\put(80,30){\circle{2}}

\put(46,4){\makebox(0,0)[t]{{\Large $p$}}}
\put(46,34){\makebox(0,0)[t]{{\Large $p$}}}
\put(70,35){\makebox(0,0)[t]{{\Large $N^*$}}}
\put(98,41){\makebox(0,0)[t]{{\Large $\eta(K^+)$}}}
\put(100,23){\makebox(0,0)[t]{{\Large $p(\Lambda)$}}}
\put(65,20){\makebox(0,0)[t]{{\Large $\pi^0$}}}
\put(75,4){\makebox(0,0)[t]{{\Large $p$}}}
\end{picture}
\caption{Feynman diagrams for the reactions $pp\rightarrow pp\eta$ and
$pp\rightarrow p\Lambda K^+$ exciting, {\it via} pion exchange, the N$^*(1535)$
and N$^*(1650)$  resonances respectively.}
\end{figure}

\newpage
\baselineskip 4ex
\begin{thebibliography}{99}
%
\bibitem{Ulf} H. Cal\'en {\it et al.}, Phys.Lett. {\bf B366} (1996) 39.
%
\bibitem{Bing} A.M. Bergdolt {\it et al.}, Phys.Rev. {\bf D48} (1993) R2969.
%
\bibitem{Beppe} E. Chiavassa {\it et al.}, Phys.Lett. {\bf B337} (1994) 192.
%
\bibitem{Ben} B.~Mayer {\it et al.}, Phys.Rev.{\bf C53} (1996) 2068.
%
\bibitem{X} J.-F. Germond and C. Wilkin, J.Phys. {\bf G15} (1989) 437, 
and Nucl.Phys. {\bf A518} (1990) 308;
T. Vetter {\it et al.}, Phys.Lett. {\bf B263} (1991) 153;\\
J.M. Laget, F. Wellers and J.F. Lecolley, Phys.Lett. {\bf B257} (1991) 254.
%
\bibitem{Stina}
S. H\"aggstr\"om, in {\it Proceedings of Workshop on Production,
Properties and Interactions of Mesons}, Cracow, Poland,
10-14 May 1996, Acta Physica Polonica, (in press);
H. Cal\'en, in {\it Proceedings of XIV International Conference
on Particles and Nuclei}, Williamsburg, Virginia, 22-28 May, 1996,
World Scientific Publishing, (in press).
%
\bibitem{Jan} J.T. Balewski {\it et al.}, Phys.Lett. {\bf B388} (1996) 859, 
and COSY experimental proposal 47 (1996).
%
\bibitem{FW1} G. F\"aldt and C. Wilkin, Nucl.Phys. {\bf A604} (1996) 441.
%
\bibitem{FW2} G. F\"aldt and C. Wilkin, Phys.Lett. {\bf B382} (1996) 209.
%
\bibitem{Paris} M.~Lacombe {\it et al.}, Phys.Rev. {\bf C21} (1980) 861.
%
\bibitem{Newton} R.G.Newton, {\it Scattering Theory of Waves and Particles}
(Springer-Verlag, N.Y., 1982).
%
\bibitem{param} B. Holzenkamp {\it et al.}, Nucl.Phys. {\bf A500} (1989) 485;\\
A. Reuber {\it et al.}, Nucl. Phys. {\bf A570} (1994) 543.
%
\bibitem{Gideon} G. Alexander {\it et al.}, Phys.Rev. {\bf 173} (1968) 1452.
%
\bibitem{Binnie} D.M. Binnie {\it et al.}, Phys.Rev. {\bf D8} (1973) 2789.
%
\bibitem{Saxon} J.J.Jones {\it et al.}, Phys.Rev.Letters {\bf 26} (1971) 860;\\
R.D. Baker {\it et al.}, Nucl.Phys. {\bf B141} (1978) 29.
\end{thebibliography}
\end{document}